\documentclass[aps,prb,twocolumn,superscriptaddress,showpacs,longbibliography]{revtex4-1}

\usepackage{amsmath,amssymb,graphicx,subfigure}

\usepackage[utf8]{inputenc}
\usepackage[T1]{fontenc}
\usepackage{xcolor}
\usepackage[normalem]{ulem}
\usepackage{multirow}
\usepackage{stmaryrd}
\usepackage{dblfloatfix} 

\IfFileExists{newtxtext.sty}
   {\usepackage{newtxtext,newtxmath}}
   {\IfFileExists{stix.sty}
      {\usepackage{stix}}
      {\IfFileExists{mathptmx.sty}
      {\usepackage{mathptmx}}{} } }

\usepackage{textcomp}

\usepackage{bm}

\IfFileExists{siunitx.sty}{\usepackage{booktabs,siunitx}}{}

\pdfoutput=1
\usepackage{color}
\definecolor{LinkColor}{rgb}{0.256,0.439,0.588}
\usepackage{hyperref}
\hypersetup{
   colorlinks=true,
   citecolor=LinkColor,
   linkcolor=LinkColor,
   urlcolor=LinkColor
}

\newcommand{\be}{\begin{equation}}
\newcommand{\ee}{\end{equation}}
\newcommand{\bea}{\begin{eqnarray}}
\newcommand{\eea}{\end{eqnarray}}

\usepackage{pifont}

\begin{document}

\title{Magnetic-field Induced Topological Transitions and Thermal Conductivity \\in a Generalized Kitaev Model}

\author{Heqiu Li}
	\affiliation{Department of Physics, University of Toronto, Toronto, Ontario M5S 1A7, Canada}
	
\author{Yong Baek Kim}
\email{ybkim@physics.utoronto.ca}
	\affiliation{Department of Physics, University of Toronto, Toronto, Ontario M5S 1A7, Canada}
	\affiliation{School of Physics, Korea Institute for Advanced Study, Seoul 02455, Korea}

\author{Hae-Young Kee}
\email{hykee@physics.utoronto.ca}
	\affiliation{Department of Physics, University of Toronto, Toronto, Ontario M5S 1A7, Canada}
	\affiliation{Canadian Institute for Advanced Research, CIFAR Program in Quantum Materials, Toronto, Ontario M5G 1M1, Canada}

\date{\today}

\begin{abstract}


Recent experiments on Kitaev spin liquid candidate materials reported non-monotonic behavior of thermal conductivity as a function of magnetic field, which lead to conflicting interpretations of its origin. Motivated by this development, we study the magnetic field dependence of thermal conductivity of a generalized Kitaev model, which allows the phase transitions between different flux sectors as a function of the magnetic field. The thermal conductivity due to Majorana fermions shows dip-bump structures as the magnetic field increases, which is caused by either the transitions between different flux sectors of Kitaev spin liquids or the topological transitions that change the Majorana Chern number within the same flux sector. It is shown that the change of Chern number is closely related to the four-Majorana-fermion interaction induced by the magnetic field. The non-monotonic behavior in thermal conductivity emerges at finite temperature, and it becomes weaker when temperature decreases toward zero. Our model provides a generic mechanism for the Kitaev spin liquids to develop non-monotonic magnetic-field dependence of thermal conductivity while the comparison to realistic materials remains an open question for future investigation.

\end{abstract}
\maketitle

\section{Introduction}

Quantum spin liquid is an exotic phase of quantum magnets that has drawn great attention since it was initially proposed~\cite{Anderson1973}. Contrary to conventional magnetic phases, quantum spin liquids are not magnetically ordered down to zero temperature, and the spins are fractionalized into spinon quasiparticles coupling to emergent gauge fields, exhibiting a lot of interesting collective phenomena including fractionalized excitations, ground state degeneracy and long range entanglement~\cite{Wen2002,Balents2010,Savary2016,Zhou2017,Hermanns2018,Knolle2019,Broholm2020}. The celebrated Kitaev honeycomb model~\cite{Kitaev2006} is exactly solvable with quantum spin liquid ground state. Kitaev spin liquids arise from the bond-dependent spin interactions that frustrate the spin orientation at each site, which can be generated from $t_{2g}$ orbitals with $j_{\rm{eff}}=\frac{1}{2}$ in materials with strong spin-orbit coupling~\cite{Jackeli2009}. The Kitaev honeycomb model provides a plausible platform for the experimental realization of quantum spin liquid, which leads to significant experimental and theoretical efforts in studying Kitaev materials~\cite{Jackeli2009,Rau2014G,Rau2014Gp,Rau2016,Trebst2017,Kasahara2018,Gohlke2018,Takagi2019,Motome2020,Ye2020,Fuchs2020,Lsz2021,Feng2021}, and many Kitaev spin liquid candidates including (Na,Li)$_2$IrO$_3$~\cite{Singh2010,Chaloupka2010,Liu2011NaIrO,Singh2012,Choi2012,Modic2014,Kim2016} and $\alpha$-RuCl$_3$~\cite{Plumb2014,Sandilands2015,Kim2015k,Sears2015,Majumder2015,Sandilands2016,Banerjee2016,Do2017,Hirobe2017} are identified.

Recent thermal transport experiments on Kitaev spin liquid candidate $\alpha$-RuCl$_3$ have reported non-monotonic dependence of longitudinal thermal conductivity on magnetic field similar to quantum oscillations~\cite{Czajka2021,Bruin2021}. There exist several explanations for the reason of this non-monotonic behavior, e.g., from the quantum oscillation from spinon Fermi surface~\cite{Czajka2021} or from phase transitions in the spin liquid~\cite{Bruin2021,Suetsugu2022}. Contrary to ordinary quantum oscillations in fermionic systems which arise from quantized Landau levels, the spinons in quantum spin liquids are charge neutral and they cannot directly couple to orbital magnetic field to give Landau levels. Therefore, the mechanism of this non-monotonic behavior in thermal conductivity remains an open question.


\begin{figure}
\includegraphics[width=3.3 in]{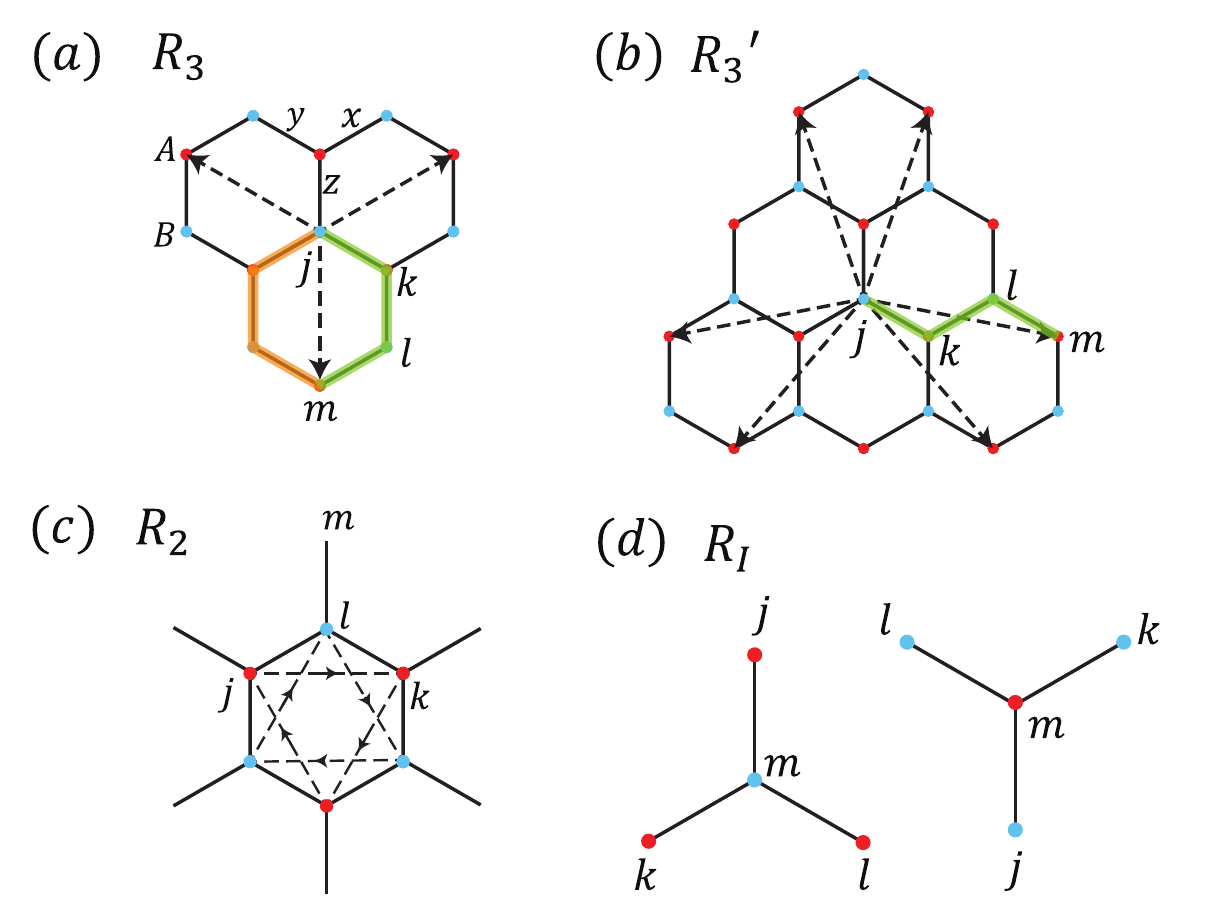}
\caption{ Illustration of terms in the Hamiltonian. A and B sublattices are labeled as red and blue dots. The dashed arrows in (a),(b),(c) represent $R_3$, $R_3'$ and $R_2$ respectively. (d) shows the relative locations of the sites in $R_I$. The sites at A and B sublattices are labeled by red and blue dots.   }
\label{K3K3p}
\end{figure}

Motivated by these experiments, in this work we explore what mechanism can lead to non-monotonic dips and bumps in thermal conductivity as a function of magnetic field. We consider a generalized Kitaev model and demonstrate that dips and bumps in thermal conductivity can arise naturally due to phase transitions induced by the magnetic field. Two important ingredients are included in this generalized Kitaev model, the frustrated further-neighbour Majorana hopping from spin interactions~\cite{Zhang2019v,Zhang2020v} and the four-Majorana interaction induced by the magnetic field.

The frustrated further-neighbour Majorana hopping allows different flux sectors to be the ground state. In Kitaev spin liquids spinons are coupled to an emergent $\mathbb Z_2$ gauge field and each plaquette can have $\mathbb Z_2$ flux $0$ or $\pi$, where $\pi$ flux corresponds to a vison. Based on the density and distribution of visons, the quantum spin liquid can be classified into different flux sectors. Each flux sector has distinct spinon spectrum and free energy. The original Kitaev honeycomb model has its ground state in the $0$-flux sector where all plaquettes have $0$ flux by Lieb's theorem~\cite{Lieb1994}. If further-neighbour Kitaev spin interactions are considered, frustrated third neighbour Majorana hopping terms~\cite{Zhang2019v,Zhang2020v} can arise. These terms commute with the $\mathbb Z_2$ gauge field hence they preserve the exact-solubility of the model. With these terms, different flux sectors can be stabilized as the ground state, which opens up the possibility for phase transitions between flux sectors. The magnetic field perturbatively generates both Majorana hopping terms and four-Majorana interaction terms. We consider both types of terms and applied self-consistent mean field theory to decouple the Majorana interaction. We found that the Majorana interaction can effectively rescale the frustrated hopping and lead to transitions that can change Majorana Chern number.

We take into account these frustrated hopping terms and the four-Majorana interaction in the generalized Kitaev model. We show that external magnetic field can lead to transitions between different flux sectors as well as topological transitions within each flux sector that change the Majorana Chern number. Because different flux sectors have distinct spinon spectrum and thermal conductivity, and because thermal conductivity is sensitive to the spinon band gap which must close when Chern number changes, these phase transitions will lead to dips and bumps in thermal conductivity as a function of magnetic field.

This paper is organized as follows. In Section \ref{Sec_model} we establish the generalized Kitaev model that includes the four-Majorana interaction terms induced by magnetic field and the frustrated Majorana hopping terms. In Section \ref{Sec_MF} we utilize self-consistent mean field theory to compute Majorana interaction. In Section \ref{Sec_thermal} we present our computation of thermal conductivity and show that it is a non-monotonic function of magnetic field with dips and bumps. In Section \ref{Sec_reason} we show that this dip-bump feature in thermal conductivity originates from phase transitions induced by magnetic field. Further relation between our results and the experiments are discussed in the discussion section.



\section{Generalized Kitaev model with multiple flux sectors}
\label{Sec_model}

The Hamiltonian of the original Kitaev honeycomb model~\cite{Kitaev2006} with nearest neighbour (NN) spin interactions is given by
\be
H_1=-K_1\sum_{\langle jk\rangle}\sigma_j^\alpha\sigma_k^\alpha=iK_1\sum_{\langle jk \rangle}u_{jk} c_jc_k,
\ee
where $j,k$ denote lattice sites, the spin operator at each site is represented by $\sigma_j^\alpha=ib_j^\alpha c_j$, and the $\mathbb{Z}_2$ gauge field $u_{j k} = i b_{j}^{\alpha} b_{k}^{\alpha}$ where $\alpha\in\{x,y,z\}$ is fixed by the direction of bond $jk$. We choose the gauge such that in the $0$-flux sector $u_{jk}=+1$ if $j$ ($k$) is at $A$ (B) sublattice, where $A,B$ sublattices are labeled in Fig.\ref{K3K3p} (a). {The original Kitaev model $H_1$ considers nearest neighbour interaction and the ground state is in the 0-flux sector. In real materials, further-neighbour spin interactions exist in general. These longer-range interactions are allowed by symmetry, and they can stabilize different flux sectors as the ground state and lead to phase transitions between different flux sectors.} Therefore, we consider a generalized Kitaev model with additional third-neighbour spin interaction terms proposed in Refs \onlinecite{Zhang2019v,Zhang2020v}:{
\bea
H_3&=&K_3 \sum_{\langle jklm \rangle\in R_3} \sigma_j^\alpha\sigma_k^\gamma\sigma_l^\alpha\sigma_m^\gamma \nonumber\\
&&-K_3' \sum_{\langle jklm \rangle\in R_3'} \sigma_j^\alpha\sigma_k^\beta\sigma_l^\beta\sigma_m^\alpha  
\eea}
Here $R_3$ and $R_3'$ denote the sets of vectors shown in Fig.\ref{K3K3p} (a) and (b) respectively, and $\langle jklm \rangle\in R_3$ means the sites $j$ and $m$ are connected by the zigzag line passing through $jklm$ with $\mathbf r_m-\mathbf r_j \in R_3$. The summation is over all such zigzag lines. $\alpha$ is along the direction of bond $jk$, $\gamma$ is along bond $lm$, and $\beta$ is distinct from the directions of $jk$ and $kl$. In terms of Majorana fermions, $H_3$ can be rewritten as
\bea
H_3&=&-iK_3 \sum_{\langle jklm \rangle\in R_3} u_{jk} u_{kl} u_{lm} c_jc_m \nonumber\\
&&-iK_3' \sum_{\langle jklm \rangle\in R_3'} u_{jk} u_{kl} u_{lm} c_jc_m \nonumber
\eea
These terms commute with gauge field $u_{jk}$, hence they preserve the exact solubility of the model and the ground state can still be labeled by flux sectors as the original Kitaev model. Note that each vector in $R_3$ corresponds to two zigzag lines (marked by colors in Fig.\ref{K3K3p} (a)), and the contribution from these two paths will cancel out if there is a $\pi$-flux inside the hexagon. Hence the $\pi$-flux sector in which every hexagon has flux $\pi$ is not affected by the $K_3$ term. It has been shown that by varying $K_3$ and $K_3'$, different flux sectors can be stabilized as the ground state~\cite{Zhang2019v,Zhang2020v}.

The external magnetic field can generate three-spin interactions in the Kitaev model as follows~\cite{Kitaev2006}:
\be
H_h={-}h \sum_{\langle jlk\rangle\in R_2}\sigma_j^\alpha\sigma_k^\beta\sigma_l^\gamma{-}h\sum_{\langle jlk\rangle\in R_I}\sigma_j^\alpha\sigma_k^\beta\sigma_l^\gamma
\ee
Here $R_2$ and $R_I$ are shown in Fig.\ref{K3K3p}(c) and (d) respectively. The choice of $\alpha,\beta,\gamma$ for the first term is along $jl,lk,lm$ respectively and for the second term is along $jm,km,lm$ respectively. These terms arise from the leading order perturbation of magnetic field and $h\sim \frac{B_xB_yB_z}{K_1^2}$. When these terms are rewritten as Majorana fermions, the first term becomes a NNN hopping term and the second term becomes a four-Majorana interaction term:
\bea
H_h&=&H_2+H_I,\ \ H_2={-}ih\sum_{\langle jlk\rangle\in R_2}u_{jl} u_{lk} c_jc_k \label{H2g}\\
H_I&=&{-}h\sum_{\Ydown}u_{jm} u_{km} u_{lm} c_mc_jc_kc_l \nonumber\\
&&{-}h\sum_{\Yup}u_{jm} u_{km} u_{lm} c_mc_jc_kc_l 
\label{HIg}
\eea
Here the relative locations of sites in $H_I$ are shown in Fig.\ref{K3K3p}(d). The total Hamiltonian for the generalized Kitaev model is
\be
H=H_1+H_2+H_3+H_I
\ee
This Hamiltonian is symmetric under space-inversion and threefold rotation. It also has a gauge symmetry indicated by $c_j\rightarrow (-1)^{n_j} c_j$, $u_{jk}\rightarrow (-1)^{n_j-n_k}u_{jk}$ where $n_j\in \{0,1\}$. Time-reversal maps $h\sim \frac{B_xB_yB_z}{K_1^2}$ to $-h$, hence time-reversal symmetry is broken by $H_2$ and $H_I$ terms induced by the magnetic field. Without the magnetic field, the Hamiltonian describes a system of free Majorana fermions for each given flux sector, and different flux sectors can be stabilized as the ground state when we vary parameters $K_3$ and $K_3'$. The magnetic field breaks time-reversal symmetry and brings in interactions between Majorana fermions. We show below that the magnetic field can induce first-order transitions between different flux sectors, and within the same flux sector the increasing magnetic field can also induce topological transitions that changes the Chern number of occupied bands. Both effects can lead to jumps and bumps in physical observables including thermal conductivity.

\section{Mean field theory for Majorana interactions}
\label{Sec_MF}

\begin{figure}
\includegraphics[width=3.4 in]{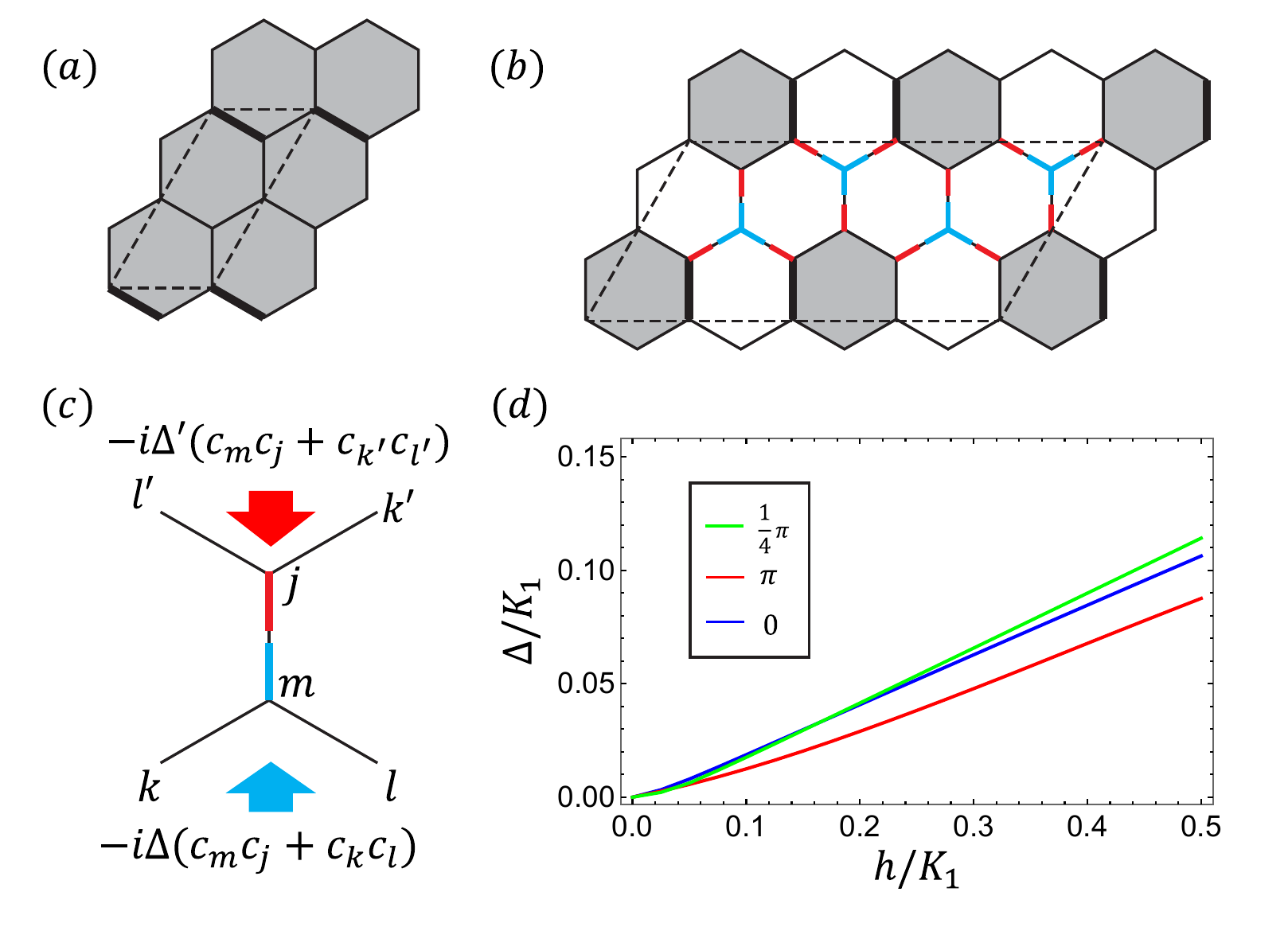}
\caption{ (a): $\pi$-flux sector in which every hexagon has flux $\pi$. The $\mathbb{Z}_2$ gauge field $u_{jk}$ flips sign at the thick black bonds to give rise to $\pi$ flux in the hexagons near it. The extended unit cell is denoted by the parallelogram. (b): $\frac{1}{4}\pi$-flux sector where white(gray) hexagons have $0$($\pi$) flux inside it. $u_{jk}$ flips sign at the thick black bonds. The extended unit cell is denoted by the parallelogram. In the mean field solution all the $\Delta$'s labeled by the blue (red) bonds have the same magnitude, which preserves the symmetry of the flux sector. (c): Different terms in the mean field Hamiltonian with $\Delta$ and $\Delta'$ are labeled by blue and red bonds respectively. (d): $\Delta$ as a function of $h$ for $0$-, $\pi$- and $\frac{1}{4}\pi$-flux sectors computed at $K_1=1, K_3=0.3, K_3'=0.34, T=\frac{1}{24}$. For $\frac{1}{4}\pi$-flux sector the $\Delta$ at the blue bonds in (b) are used. $\Delta$ increases with $h$ as magnetic field increases, and the $\Delta$ becomes linear in $h$ when $h$ is large.       }
\label{figdelta}
\end{figure}

We handle the four-Majorana interactions $H_I$ in Eq.\eqref{HIg} by self-consistent mean field theory. For each flux sector, $H_I$ contains terms like $\pm |h| c_mc_jc_kc_l$. Denote the center of each $\Yup$ or $\Ydown$ by site $m$. Note that $\pm|h|  c_mc_jc_kc_l=-\frac{|h|}{2}(\pm ic_mc_j+ic_kc_l)^2+|h|$, we can make a Hubbard-Stratonovich (H-S) transformation to decouple the square term. To preserve the threefold rotational symmetry we introduce three real auxiliary fields $\Delta_{jk},\Delta_{kl},\Delta_{lj}$. {Define $\eta_{mjkl}=-\text{sign}(h)u_{jm} u_{km} u_{lm}=\pm 1$, then the H-S transformation is given by:
\bea
&&-h u_{jm} u_{km} u_{lm} c_mc_jc_kc_l=|h|\eta_{mjkl}c_mc_jc_kc_l \nonumber\\
&&\rightarrow \frac{3\Delta_{kl}^2}{2|h|}-i\Delta_{kl}(\eta_{mjkl}c_mc_j+c_kc_l) \nonumber \\
&&\ \ \ +\frac{3\Delta_{lj}^2}{2|h|}-i\Delta_{lj}(\eta_{mjkl}c_mc_k+c_lc_j)\nonumber \\
&&\ \ \ +\frac{3\Delta_{jk}^2}{2|h|}-i\Delta_{jk}(\eta_{mjkl}c_mc_l+c_jc_k)+|h|
\label{HStransform}
\eea
Here the factor of 3 is to ensure integrating out the $\Delta$'s recovers the original four-fermion interaction, and $\eta_{mjkl}$ keeps track of the overall sign of the four-Majorana term.} We apply Eq.\eqref{HStransform} to $H_I$ with distinct $\Delta$'s for each term, then we have decoupled the Majorana interaction into free Majorana coupled to auxiliary fields. We further require the terms related by a translation of (extended) lattice vector have the same $\Delta$, then the decoupling preserves the translational symmetry. The number of independent $\Delta$'s is three times the number of sites in the extended unit cell for each flux sector. For example, the 0-flux sector has two sites in the unit cell and it needs six $\Delta$'s. The $\pi$-flux sector is shown in Fig.\ref{figdelta}(a), where the signs of the $\mathbb{Z}_2$ gauge field $u_{jk}$ at the thick black bonds are flipped. The extended unit cell of the $\pi$-flux sector has four sites and it needs twelve $\Delta$'s. 

The mean field solution is obtained by finding the set of $\Delta$'s that minimizes the free energy. This is also equivalent to finding $\Delta$'s that satisfy the self-consistency equations:{
\bea
\Delta_{kl}&=&\frac{|h|}{3}\langle i\eta_{mjkl}c_mc_j+ic_kc_l  \rangle \nonumber \\
\Delta_{lj}&=&\frac{|h|}{3}\langle i\eta_{mjkl}c_mc_k+ic_lc_j  \rangle \nonumber \\
\Delta_{jk}&=&\frac{|h|}{3}\langle i\eta_{mjkl}c_mc_l+ic_jc_k  \rangle 
\label{selfc}
\eea
Here the average denotes the thermal average at the same temperature as the temperature under which the free energy is computed.} It is evident from Eq.\eqref{selfc} that the $\Delta$'s are real. Our mean field theory is different from Ref~\onlinecite{Majohubbard2018} in that our decoupling ensures the mean field solution obtained by solving the self-consistency equations and minimizing free energy agree with each other. We have computed the mean field solutions for different flux sectors with flux $0,\pi,\frac{1}{2}\pi,\frac{1}{3}\pi,\frac{2}{3}\pi,\frac{1}{4}\pi,\frac{3}{4}\pi$. 

Our solutions show that the $\Delta$'s respect the symmetries of each flux sector. For example, the $0$-flux sector has uniform flux distribution in space, and all the $\Delta$'s in Eq.\eqref{HStransform} have the same magnitude in the mean field solution. We can label each $\Delta$ term by a colored bond with half the length of a NN bond as in Fig.\ref{figdelta}(c). The sign of $\Delta$'s are distributed in such a way that it modifies the NN hopping strength from $K_1$ to $K_1+2\Delta$, where the factor of 2 is because each NN bond is shared by two $\Delta$'s as in Fig.\ref{figdelta}(c), and the NNN hopping strength is modified from $h$ to $h+\Delta$. The uniform magnitude of $\Delta$'s also appears in the $\pi$-flux sector. For other flux sectors whose flux distribution in space is not uniform, e.g., the $\frac{1}{4}\pi$-flux phase in Fig.\ref{figdelta}(b), only the symmetry-related $\Delta$'s have the same magnitude. Here all the $\Delta$'s at the blue bonds in Fig.\ref{figdelta}(b) have the same value $\Delta_a$, and all $\Delta$'s at the red bonds have another value $\Delta_b$. As the magnetic field increases, $\Delta$ also increases and the ratio $\Delta/h$ becomes a constant at large $h$, as shown Fig.\ref{figdelta}(d). With this mean field solution, we are able to compute the Majorana spectrum and the evolution of thermal conductivity with magnetic field.

\section{Thermal conductivity}
\label{Sec_thermal}

Thermal conductivity can be computed from the mean field Hamiltonian for each flux sector. The Hamiltonian in momentum space is $H=\sum_\mathbf{k} \psi_{\mathbf{k},i}^\dagger H^\Delta(\mathbf{k})_{ij}\psi_{\mathbf{k},j} $, where $\psi_{\mathbf{k},j}=\frac{1}{{\sqrt{2N}}}\sum_j e^{i\mathbf{k}\cdot \mathbf{r}_j}c_j$ is the Fourier transform of Majorana operators, $H^\Delta(\mathbf{k})$ is the Hamiltonian matrix that also depends on $\Delta$'s. The system has zero chemical potential, and the heat current operator is given by~\cite{Katsura2010}
\be
\mathbf{J}^Q=\sum_\mathbf{k} \psi_{\mathbf{k},i}^\dagger \left(\frac{1}{2} \partial_{\mathbf{k}} (H^\Delta(\mathbf{k})^2)\right)_{ij}\psi_{\mathbf{k},j}
\ee
The current-current correlation in Matsubara frequency is
\be
\Pi_{\mu\nu}(i\Omega_n)=-\int_0^\beta d\tau e^{i\Omega_n \tau} \langle T_\tau J_\mu^{Q}(\tau)J_\nu^Q(0)  \rangle
\ee
The thermal conductivity $\kappa_{xx}$ can be computed by~\cite{Mahan2013book,Durst2000,Wang2021magnon}
\bea
\kappa_{xx}&=&-\lim_{\Omega\rightarrow 0}\frac{\text{Im }\Pi_{xx}(\Omega+i\delta)}{\Omega T} \nonumber\\
&=&\lim_{\Omega\rightarrow 0}\pi\int \frac{d^2k}{(2\pi)^2}\int_{-\infty}^{\infty} d\omega \frac{n_F(\omega)-n_F(\omega+\Omega)}{\Omega T} \nonumber\\
&&\ \  \text{Tr}\left[ J_x^Q(\mathbf k) A(\mathbf{k},\omega) J_x^Q(\mathbf k) A(\mathbf k,\omega+\Omega) \right], 
\label{Kxxeq}
\eea
where $n_F$ is the Fermi distribution function, $J_x^Q(\mathbf k)$ is the matrix of heat current operator, $A(\mathbf k,\omega)=-\frac{1}{\pi}\text{Im}G^{\text{ret}}(\mathbf k,\omega)=\frac{\gamma}{(\omega-H^\Delta(\mathbf k))^2+\gamma^2}$ is the matrix of spectral function and we have added a small impurity scattering rate {$\gamma=0.03K_1$}. Eq.\eqref{Kxxeq} involves the derivative of $n_F$ which is peaked at the Fermi energy with a width proportional to temperature $T$, hence only the states with energy close to the Fermi level within the scale of $T$ can contribute to $\kappa_{xx}$. This indicates thermal conductivity will be small if the density of states near the Fermi energy is low. 

\begin{figure}
\includegraphics[width=3.4 in]{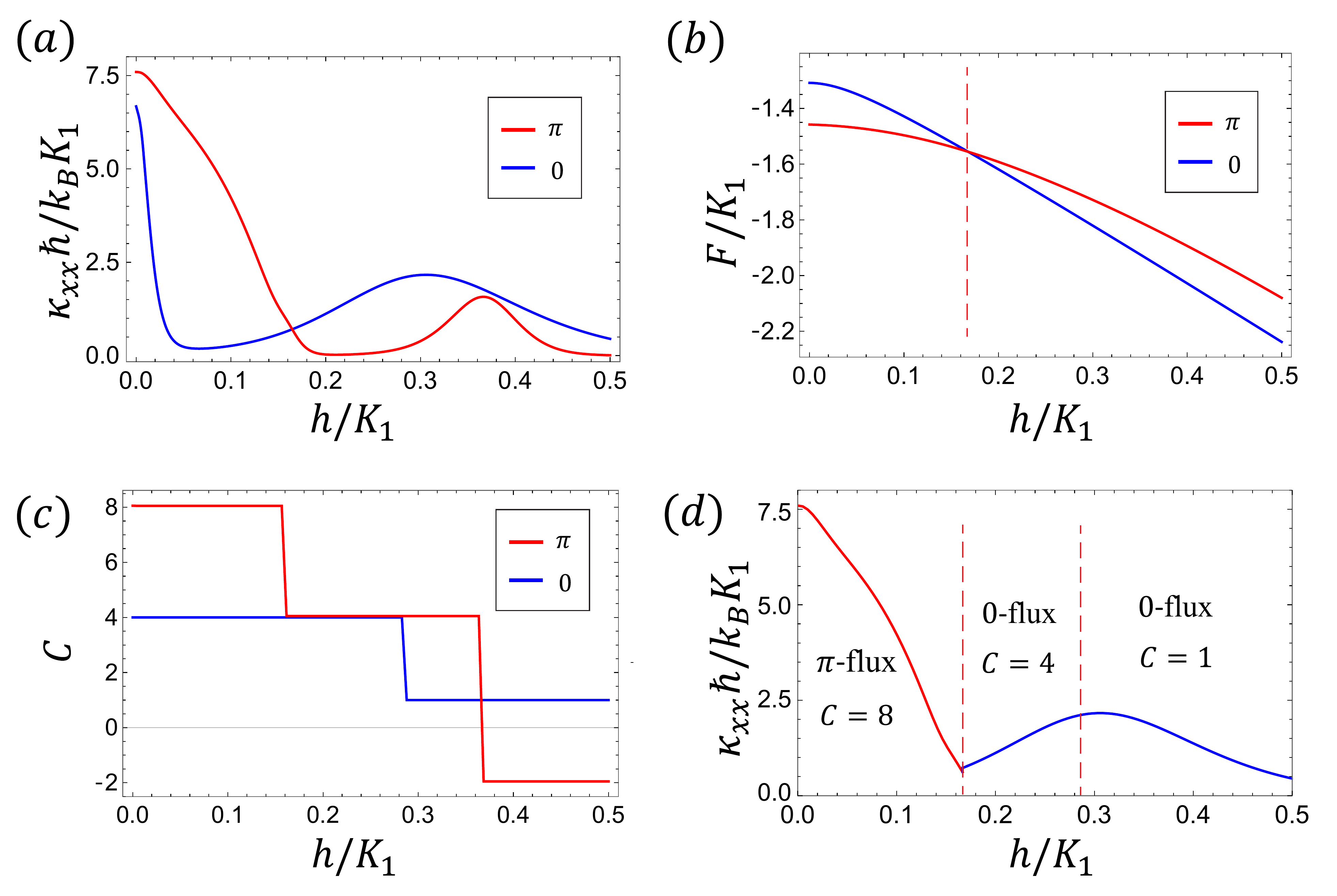}
\caption{ Physical quantities as a function of $h/K_1$ for $0$- and $\pi$-flux sectors with parameters $K_1=1,K_3=0.3,K_3'=0.34,T=\frac{1}{24}$. We only show the results for these two flux sectors because they have lower free energy than other flux sectors. (a) Thermal conductivity. (b) Free energy. (c) Chern number. (d) Thermal conductivity of the flux sector with the lowest free energy. The maxima of $\kappa_{xx}$ in (a) coincide with the changes of Chern number in (c). The red dashed lines in (d) represent the transition from $\pi$- to $0$-flux sector and the topological transition that changes Chern number. These transitions give rise to dip-bump features in thermal conductivity as a function of magnetic field.  }
\label{Kxxplot}
\end{figure}

\begin{figure}
\includegraphics[width=3.4 in]{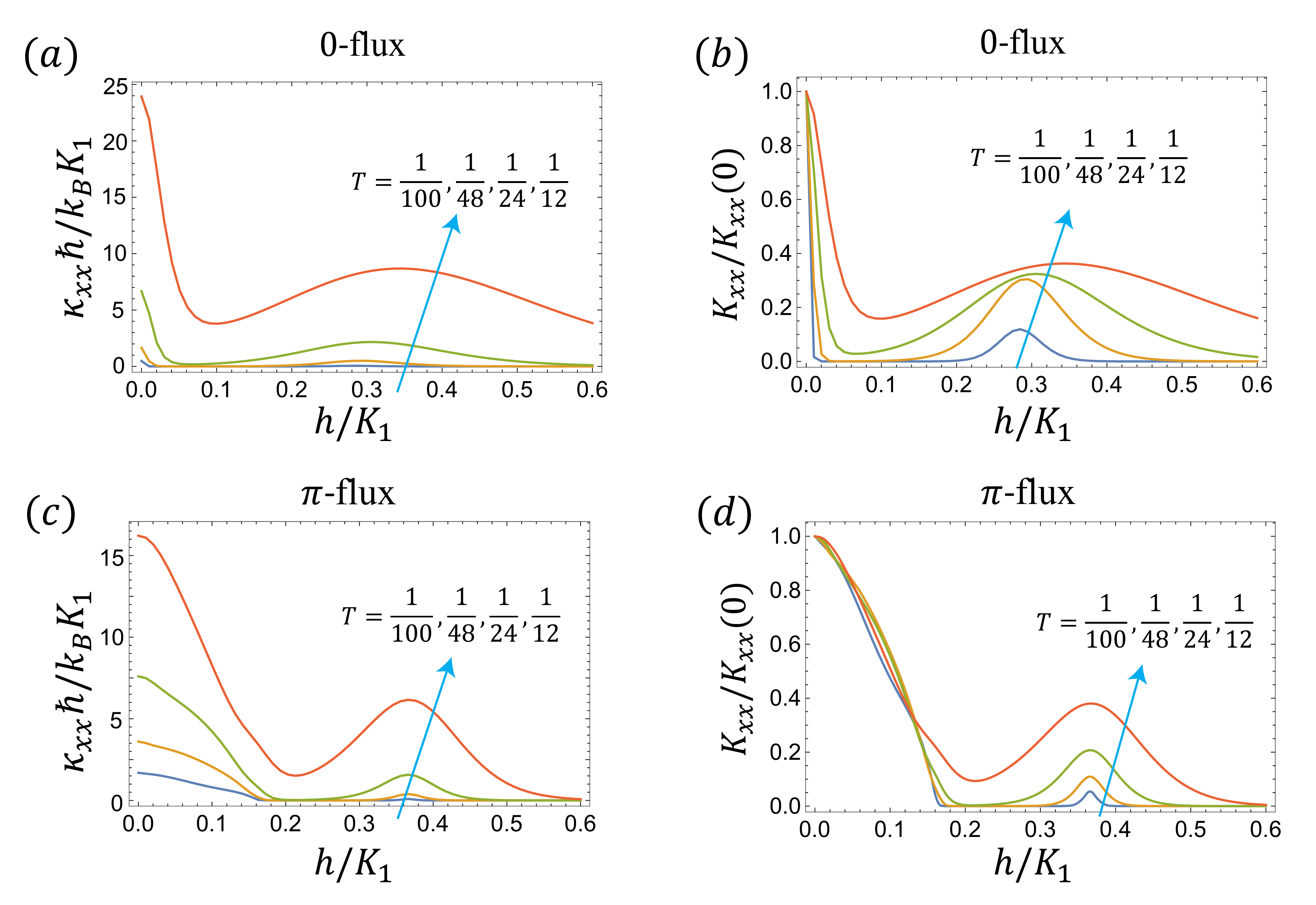}
\caption{ Thermal conductivity as a function of $h/K_1$ for $0$- and $\pi$-flux sectors at different temperature with $K_1=1,K_3=0.3,K_3'=0.34$. (a), (b) are for $0$-flux sector and (c), (d) are for $\pi$-flux sector. (b) and (d) are rescaled by $\kappa_{xx}$ at $h=0$. The dips and bumps become weaker at low temperature.   }
\label{KxxT}
\end{figure}

The thermal conductivity as a function of $h/K_1$ for $0$- and $\pi$-flux sectors with parameters $K_1=1,K_3=0.3,K_3'=0.34,T=\frac{1}{24}$ are plotted in Fig.\ref{Kxxplot}(a). We only show these two flux sectors because they have lower free energy than the other flux sectors, and the results for the other flux sectors are shown in the appendix. Fig.\ref{Kxxplot}(a) shows that the $\kappa_{xx}$ curves develop dips and bumps as a function of magnetic field. The detailed shape of the curves depend on the parameters, but the appearance of dip-bump features in each flux sector is generic. The free energy with the same parameters is shown in Fig.\ref{Kxxplot}(b). It shows the ground state is in the $\pi$-flux sector for small $h$ and it goes through a transition to the $0$-flux sector at large $h$. This first-order transition can lead to a jump in physical observables. Fig.\ref{Kxxplot}(d) shows the thermal conductivity of the flux sector with the lowest free energy. The red dashed line represents the phase transition from the $\pi$- to $0$-flux sector, which is accompanied by a dip in $\kappa_{xx}$. The temperature dependence of thermal conductivity for $0$- and $\pi$ flux sectors are shown in Fig.\ref{KxxT}, where in (b) and (d) the curves are normalized by $\kappa_{xx}$ at $h=0$. It shows that the dips and bumps become weaker at low temperature, because thermal conductivity is a transport property at finite temperature which will decrease when temperature is low.


\section{Dips and bumps in thermal conductivity induced by phase transitions}
\label{Sec_reason}

\begin{figure}
\includegraphics[width=3.4 in]{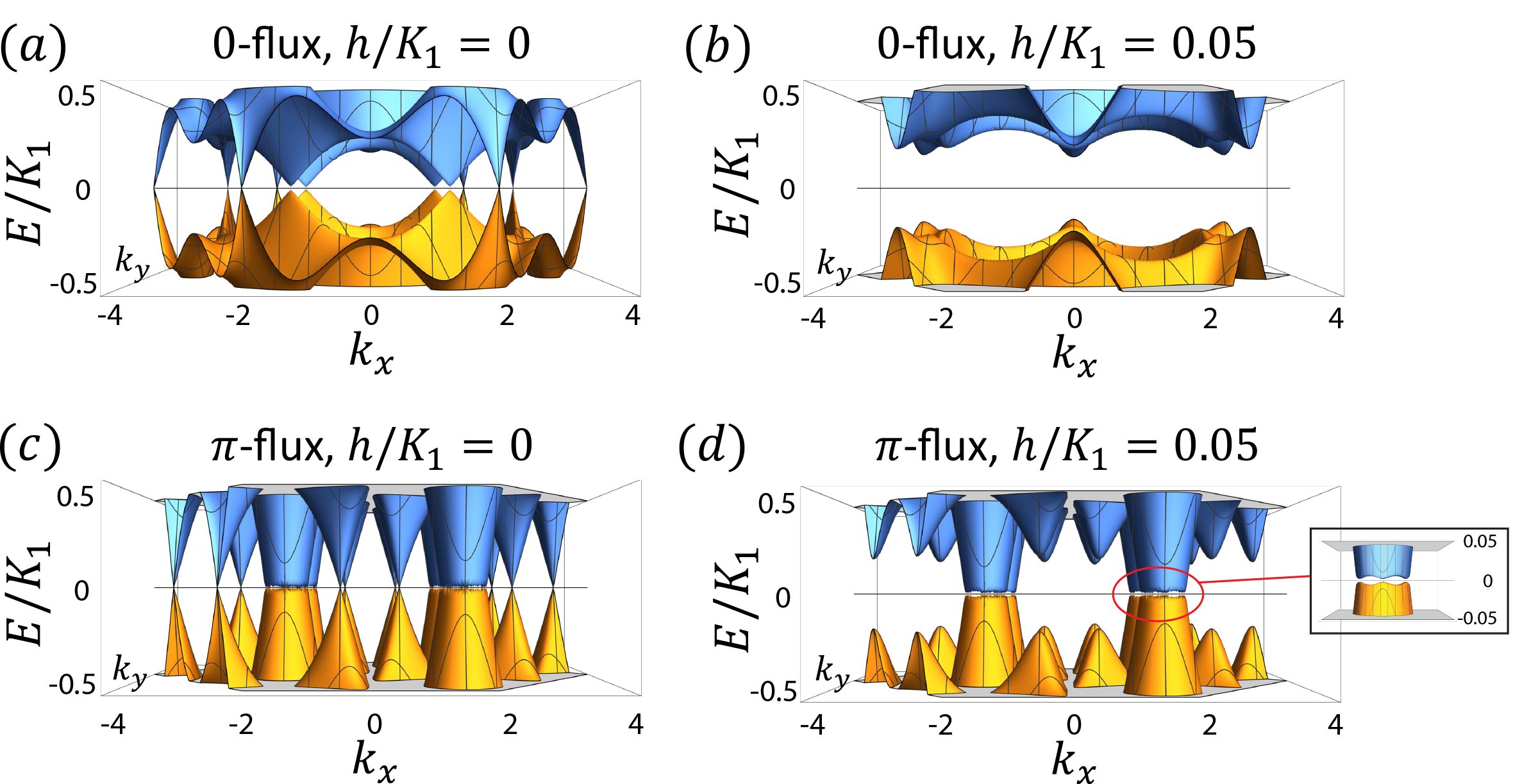}
\caption{ Dispersion of Majorana bands near the Fermi energy for $0$- and $\pi$- flux sectors with $K_1=1, K_3=0.3, K_3'=0.34$. For both flux sectors, the spectrum are gapless at $h=0$, and a small $h$ induced by magnetic field can open a gap. The gap in the $\pi$-flux sector is much smaller than that for the $0$-flux sector.     }
\label{figEk0pi}
\end{figure}

The dips and bumps of thermal conductivity as a function of magnetic field shown in Fig.\ref{Kxxplot}(a) and (d) is a generic feature that exists in a wide range in the parameter space. These phenomena can be understood from the transition between different flux sectors and the gap closing accompanied by the topological transitions that changes the Chern number. When the magnetic field is small, it tends to open a gap in the Majorana fermion bands for $0$- and $\pi$-flux sectors. The low energy band structure $E(\mathbf k)$ of $0$- and $\pi$-flux sectors for the parameters in Fig.\ref{Kxxplot}(a) at zero and small $h$ are shown in Fig.\ref{figEk0pi}. The spectrum at $h=0$ without magnetic field is gapless for both flux sectors. When a small magnetic field is added, the spectrum in both flux sectors open a gap. From Eq.\eqref{Kxxeq} the thermal conductivity is related to the density of states near the Fermi energy, therefore this gap reduces the thermal conductivity, leading to the decrease of $\kappa_{xx}$ at small $h$ for $0$- and $\pi$-flux sectors in Fig.\ref{Kxxplot}(a). 


When magnetic field increases, there is a transition from $\pi$- to $0$-flux sector, which leads to a dip in thermal conductivity at $h/K_1=0.165$. As the magnetic field further increases, it can induce topological transitions that change the Chern number for the Majorana bands. The Chern number for each flux sector is shown in Fig.\ref{Kxxplot}(c), and a comparison with Fig.\ref{Kxxplot}(a) shows that the local maxima of thermal conductivity coincide with the change in Chern number. This is because the gap needs to close near transitions that change the Chern number, which leads to an increase in the density of states near Fermi energy. Because $\kappa_{xx}$ is related to the density of state, this change in Chern number leads to an increase of $\kappa_{xx}$, which is the origin of the bump of thermal conductivity near $h/K_1=0.3$.


\begin{figure}
\includegraphics[width=2.4 in]{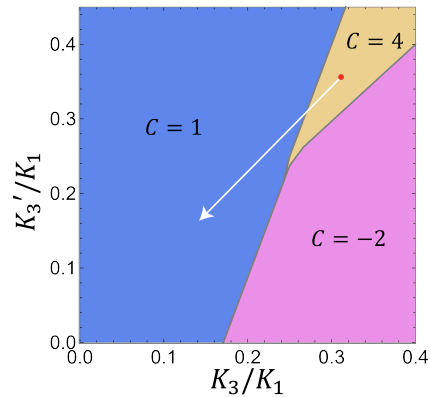}
\caption{ Chern number of the $0$-flux sector with finite $H_2$ and vanishing $H_I$. This phase diagram is independent of the magnitude of $H_2$ as long as $H_2$ is finite. Interaction $H_I$ induced by the magnetic field can rescale the effective $K_3$ and $K_3'$  along the white arrow, leading to a phase transition that changes the Chern number.   }
\label{Chern0}
\end{figure}

The Majorana interaction term $H_I$ induced by the magnetic field also plays an important role in the change of Chern number. Consider the $0$-flux sector for simplicity in which all the $\Delta$'s have the same value in the mean field solution. If we turn off interaction $H_I$ but keep a finite $H_2$ (see Eq.\eqref{H2g}), the phase diagram of Chern number for the $0$-flux sector with different $K_3$ and $K_3'$ is shown in Fig.\ref{Chern0}. From Eq.\eqref{HStransform} the auxiliary field $\Delta$ modifies the NN hopping terms in the mean field Hamiltonian by changing $K_1$ to $K_1+2\Delta$. Because an overall scaling of Hamiltonian does not modify the Chern number, a system with parameters $\{K_1+2\Delta,K_3,K_3'\}$ has the same Chern number as the one with parameters $\{K_1,K_3\frac{K_1}{K_1+2\Delta},K_3'\frac{K_1}{K_1+2\Delta}\}$. Therefore, the interaction term effectively rescales $K_3$ and $K_3'$ by a factor $\frac{K_1}{K_1+2\Delta}$. The mean field solution shows $\Delta$ is a positive number that grows with $h$. Therefore, if the system without magnetic field has $K_3,K_3'$ located at, e.g., the red dot in Fig.\ref{Chern0}, then as the magnetic field increases, $h$ and $\Delta$ will increase and the effective $K_3$ and $K_3'$ will follow the white arrow in Fig.\ref{Chern0}. When the effective $K_3$ and $K_3'$ cross a phase boundary between different Chern numbers, the Chern number will change and the gap will close, leading to an increase in $\kappa_{xx}$. This explains the bump of $\kappa_{xx}$ in the $0$-flux sector in Fig.\ref{Kxxplot}(a) and (d) at $h/K_1=0.3$. As long as $K_3$ and $K_3'$ are chosen to be at $C=4$ or $C=-2$ region close to the phase boundary, the bump in $K_{xx}$ is expected to occur, which is not sensitive to the exact choice of $K_3$ and $K_3'$. Therefore, the change of Chern number induced by magnetic field is a robust feature of this Majorana system which is independent of the detailed parameter choice.

\section{Discussions}

We have shown that in quantum spin liquids an external magnetic field can induce dips and bumps in thermal conductivity $\kappa_{xx}$ via phase transitions between different flux sectors and different Chern numbers. It is tempting to compare our results to the recent experiments where oscillations of $\kappa_{xx}$ with magnetic field was observed. The dip-bump features that we found in $\kappa_{xx}$ become weaker at lower temperature, as shown in Fig.\ref{KxxT}, which could be consistent with the report in Ref~\onlinecite{Czajka2021} of a weakening trend of the oscillatory features at lower temperatures. Our model is a possible mechanism for the non-monotonic behavior in thermal conductivity, but it is difficult for a direct comparison between our model and realistic materials. Our model predicts the change in Chern number accompanies the local maxima of thermal conductivity, which will lead to a jump in thermal Hall conductivity $\kappa_{xy}$ because $\kappa_{xy}$ is related to Chern number, and this change in $\kappa_{xy}$ will be sharper at lower temperature. The change of $\kappa_{xy}$ near the bump of $\kappa_{xx}$ has not been reported in candidate materials. The relation between $\kappa_{xy}$ and the non-monotonic behavior of $\kappa_{xx}$ are left as an open question for future investigation.

\section{Acknowledgment}

This work is supported by the Natural Sciences and Engineering Research Council of Canada (NSERC) and the Center for Quantum Materials at the University of Toronto. H.Y.K acknowledges the support by the Canadian Institute for Advanced Research (CIFAR) and the Canada Research Chairs Program. Computations were performed on the Niagara supercomputer at the SciNet HPC Consortium. SciNet is funded by: the Canada Foundation for Innovation under the auspices of Compute Canada; the Government of Ontario; Ontario Research Fund - Research Excellence; and the University of Toronto.


\appendix

\setcounter{equation}{0}
\setcounter{figure}{0}
\setcounter{table}{0}
\makeatletter
\renewcommand{\theequation}{S\arabic{equation}}
\renewcommand{\thefigure}{S\arabic{figure}}

\section{Thermal conductivity in other flux sectors}

\begin{widetext}

\begin{figure}[h]
\includegraphics[width=6.8 in]{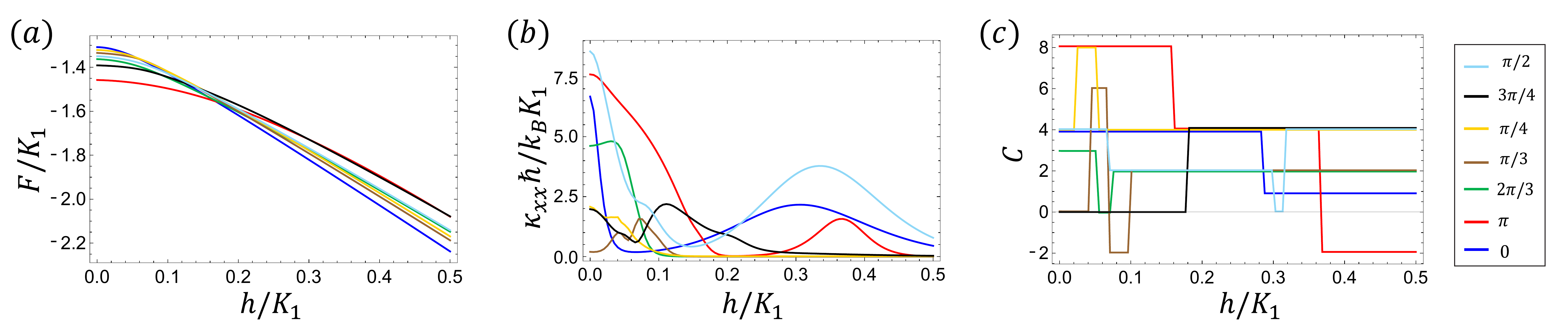}
\caption{ Free energy (a), thermal conductivity (b) and Chern number (c) as a function of $h$ induced by magnetic field for various flux sectors with $K_1=1, K_3=0.3, K_3'=0.34, T=\frac{1}{24}$. In (c) the overlapped curves are slightly shifted for better visibility.   }
\label{app_plot}
\end{figure}

\end{widetext}

We have computed the thermal conductivity for various flux sectors including $0, \pi,\frac{1}{2}\pi,\frac{1}{3}\pi,\frac{2}{3}\pi,\frac{1}{4}\pi$, and $\frac{3}{4}\pi$. The plots for $0$- and $\pi$-flux sectors are shown in the main text. Here we show the results for all these flux sectors in Fig.\ref{app_plot}. The free energy for each flux sector as a function of $h$ induced by magnetic field is plotted in Fig.\ref{app_plot}(a). It shows the ground state with the lowest free energy is in $0$- or $\pi$-flux sector. The thermal conductivity $\kappa_{xx}$ is plotted in Fig.\ref{app_plot}(b), which shows the dependence of $\kappa_{xx}$ on magnetic field is non-monotonic in general within each flux sector. This non-monotonic behaviour is related to the change in Chern number as shown in Fig.\ref{app_plot}(c). Within each flux sector, each local maximum of $\kappa_{xx}$ coincides with a topological transition that changes the Chern number. This is because the change in Chern number accompanies the closure of Majorana band gap and an increase in density of states, which leads to increased thermal conductivity. Since the transitions that change the Chern number within each flux sector generically exist, and the transitions between different flux sectors driven by free energy can also lead to jumps in thermal conductivity, these transitions provide a generic mechanism to induce non-monotonic dependence of thermal conductivity on magnetic field.

%

\end{document}